\documentclass[12pt]{article}

\usepackage{graphicx,epsfig}

\usepackage{scicite}
\usepackage{times}
\newenvironment{sciabstract}{%
\begin{quote} \bf}
{\end{quote}}

\newcommand{\vk}{{\mathbf{k}}}

\newcounter{lastnote}

\title{ Modelling the Localized to Itinerant Electronic
  Transition in the Heavy Fermion System CeIrIn$_5$ }

\author
{J.H. Shim, K. Haule and G. Kotliar\\
\\
\normalsize{Center for  Materials Theory, Department of Physics
and Astronomy,} \\
\normalsize{Rutgers University, Piscataway, NJ 08854}
}

\date{}

\begin{document} 

\baselineskip24pt

\maketitle

\begin{sciabstract}
We address the fundamental question of crossover from localized to
itinerant state of a paradigmatic heavy fermion material CeIrIn$_5$.
The temperature evolution of the one electron spectra and the optical
conductivity is predicted from first principles calculation.
The buildup of coherence in the form of a dispersive many body feature
is followed in detail and its effects on the conduction electrons
of the material is revealed.  We find multiple hybridization gaps and
link them to the crystal structure of the material. Our theoretical
approach explains the multiple peak structures observed in optical
experiments and the sensitivity of CeIrIn$_5$ to substitutions of the
transition metal element and may provide a microscopic basis for the
more phenomenological descriptions currently used to interpret
experiments in heavy fermion systems.
\end{sciabstract}


Heavy fermion materials have unusual properties arising from the
presence of a partially filled shell of $f$-orbitals and a very broad
band of conduction electrons. At high temperatures, the $f$-electrons
behave as atomic local moments. As the temperature is reduced, the
moments combine with the conduction electrons to form a fluid of very
heavy quasiparticles, with masses which are two to three orders of
magnitude larger than the mass of the electron \cite{Stewart,Allen}.

These heavy quasiparticles can undergo superconducting or magnetic
transitions at much lower temperatures. Understanding how the
itinerant low energy excitations emerge from the localized moments of
the $f$ shell is one of the central challenges of the condensed matter
physics.  It requires the understanding of how the dual, atomic
particle-like and itinerant wave-like, character of the electron
manifests itself in the different physical properties of a material.

\begin{figure}
\includegraphics[width=1.0\linewidth]{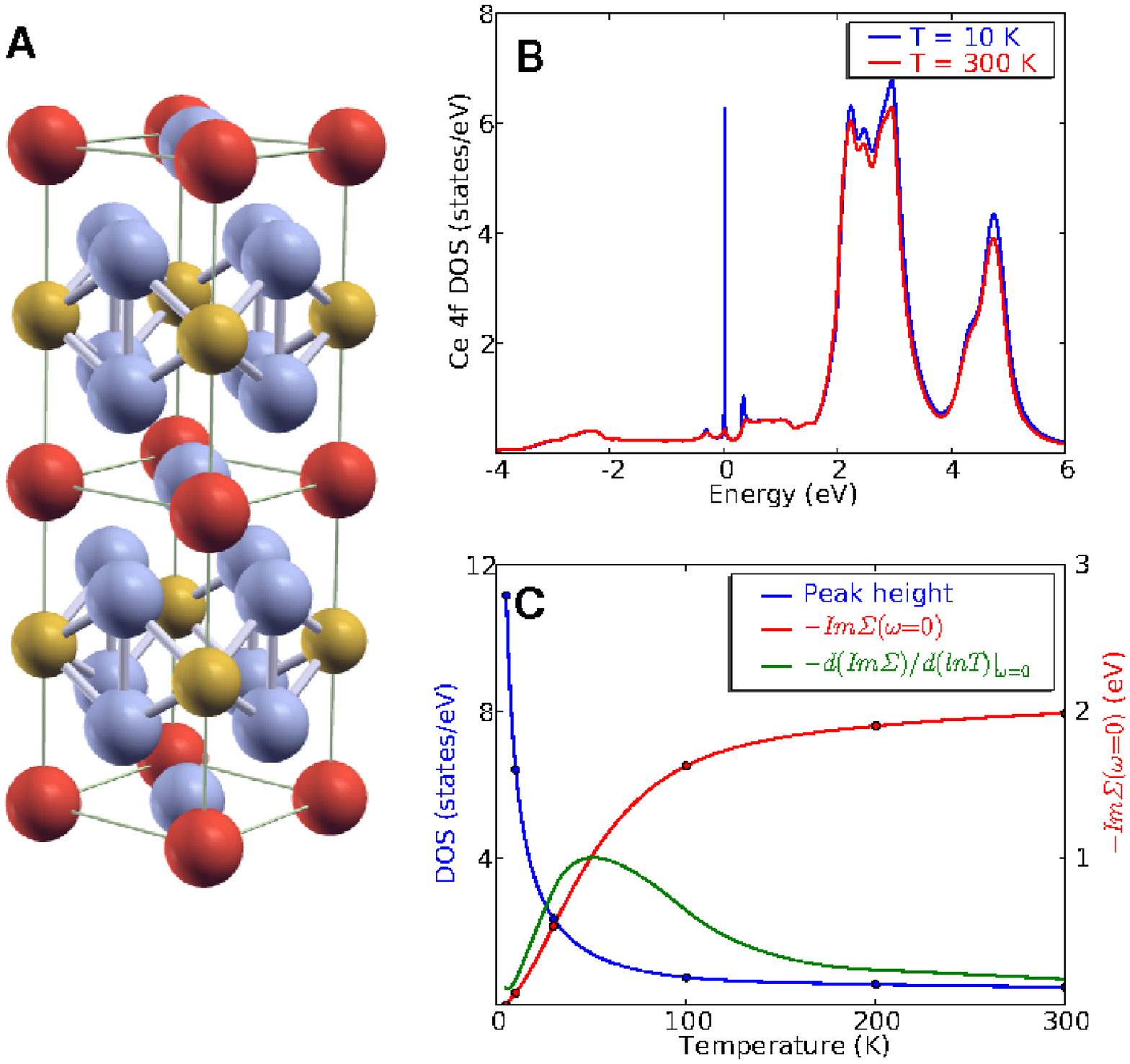}
\caption{
(A) Crystal structure of CeIrIn$_5$. Red, gold, and gray spheres
correspond Ce, Ir, and In atoms, respectively.
(B) Ce $4f$ local density of state calculated by LDA+DMFT at $10$~K and $300$~K.
(C) The quasiparticle peak hight versus temperature (blue), imaginary
part of Ce $4f_{5/2}$ self-energy $\Sigma_f(\omega=0)$(red) and its
temperature derivative (green).  The buildup of coherence is very slow
and gradual. Around $T^*\sim 50~K$ the coherence first sets in and
manifested itself in fast increase of the quasiparticle peak (blue line) and
crossover in the scattering rate (the derivative of scattering rate -
green line - is peaked at $T^*$). The quasiparticle peak weight
saturates at much lower temperature $<5K$ and drops to zero at very
high temperature, displaying a very long logarithmic tail.
}
\label{fig1}
\end{figure}

CeIrIn$_5$ \cite{Petrovic1} has a layered tetragonal crystal structure
\cite{Grin,Moshopoulou} (Fig.~\ref{fig1}A) in which the layers of CeIn$_3$
(shown as red and gray spheres) are stacked between layers of IrIn$_2$
(yellow and grey spheres). Each Ce atom is surrounded by four In atoms
in the same plane and eight In atoms out of plane.

To describe the electronic structure of these class of materials, one
needs to go beyond the traditional concepts of bands and atomic
states, and focus on the concept of a spectral function
$A(\vk,\omega)_{LL}$, which describes the quantum mechanical
probability of removing or adding an electron with angular momentum
and atomic character $L=(l,m,a)$, momentum $\vk$ and energy
$\omega$. It is measured directly in angle resolved photoemission and
inverse photoemission experiments.

To evaluate the spectral function we use Dynamical Mean Field Theory
(DMFT) \cite{GabiDieter} in combination with the Local Density
Approximation (LDA+DMFT)\cite{Kotliar}, which can treat the realistic
band structure, the atomic multiplet splitting and Kondo screening on
the same footing.  The spectral function is computed from
corresponding one-electron Green's function
$A(\vk,\omega)=(G^\dagger(\vk,\omega)-G(\vk,\omega))/(2\pi i)$ where
the latter takes the form
\begin{equation}
  G(\vk,\omega) = \frac{1}{(\omega+\mu)O_\vk-H_\vk-\Sigma(\omega)}.
\end{equation}
Here $H_\vk$ and $O_\vk$ are the Hamiltonian and overlap matrix
obtained by the LDA method\cite{Savrasov} and $\Sigma$ is the DMFT
self-energy, which requires a solution of the quantum impurity problem
embedded in a self consistent medium. We used a
vertex corrected one-crossing approximation\cite{Kotliar} and the results
were further crosschecked against a continuous time quantum Monte
Carlo method\cite{Werner,Haule}.
The Slater integrals, F2,F4,F6, were computed by the atomic physics
program of Ref.~\citen{Cowan} and F0 was estimated by the constrained LDA to be
5~eV\cite{Ce_U_5}. The localized Ce-$4f$-orbital
was constructed from the non-orthogonal Linear-Muffin-Tin-orbitals in
particular way to maximize its $f$ character, as explained elsewhere\cite{Toropova}.


The spectral function of $f$ electron materials has been known to
exhibit remarkable many body effects. To set the stage for their
theoretical description, Fig.~\ref{fig1}B displays the Ce-$4f$ local
spectral function, i.e., $A(\omega) = \sum_\vk A(\vk,\omega)$, which
is measured in angle integrated photoemission experiments.

At room temperature, there is very little spectral weight at the Fermi
level as the $f$ electrons are tightly bound and localized on the Ce
atom, giving rise to a broad spectrum concentrated mainly in the lower
and upper Hubbard bands at -2.5eV and +3eV, respectively.

As the temperature is decreased, a narrow peak appears near the Fermi
level (see Fig.~\ref{fig1}B). The
states forming this peak have a small but finite dispersion, and
therefore the area of the peak can be interpreted as the degree of $f$
electron delocalization. This quantity as well as the scattering rate
of the Ce-$4f$ states ($\textrm{Im}\Sigma(\omega=0)$) exhibit a clear
crossover at a temperature scale $T^*$ of the order of 50~K (Fig.~\ref{fig1}C).

Our results are consistent with the angle integrated photoemission
measurements \cite{Fujimori-2006} in which the
onset of states with $f$ character at the Fermi level were observed. But the
experimental resolution has to be improved by one order of magnitude
to resolve the narrow peak predicted by the theory.

\begin{figure}
\includegraphics[width=0.9\linewidth]{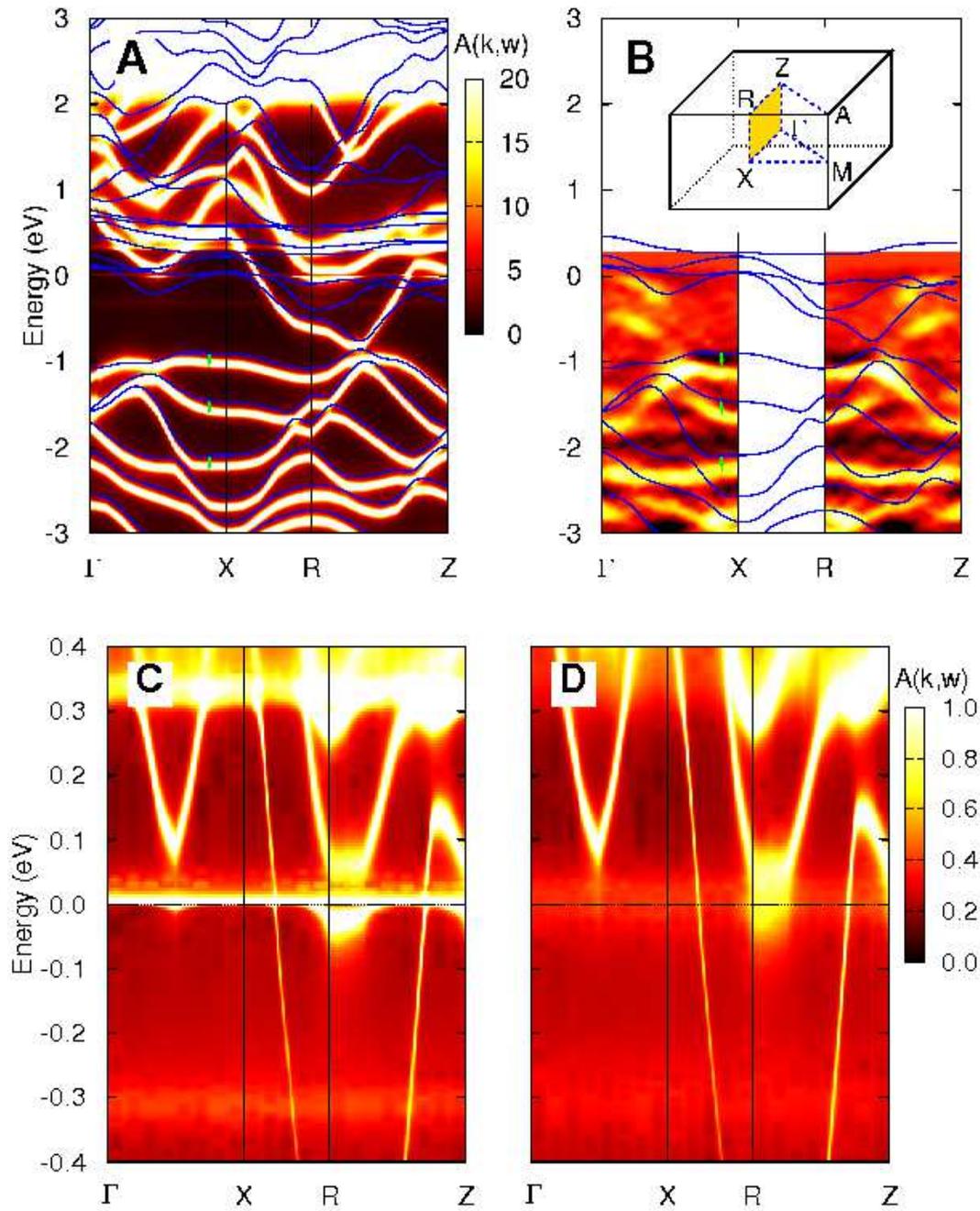}
\caption{
(A) Momentum resolved total spectral functions calculated by LDA+DMFT
method at 10~K is shown by color scheme. The LDA bands are drawn by
blue lines.
(B) Color plot shows experimental ARPES data reproduced from
Ref.~\citen{Fujimori-2003}. Blue lines follow the LDA bands.
(C) Momentum resolved Ce-$4f$ spectral function at 10~K.
(D) same as (C) but at 300~K.
}
\label{fig2}
\end{figure}

We now turn to the total (traced over all orbitals) momentum resolved
spectral function $\textrm{Tr}[A(\vk,\omega)]$ plotted along symmetry
directions in the Brillouin zone. In a band theory description, it
would be sharply peaked on a series of bands $\epsilon_n(\vk)$ and the
weight of those peaks would be unity.
It is worthwhile comparing the high intensity features of the LDA+DMFT
spectra (color coded) with the LDA bands ($\epsilon_n(\vk)$ - drawn in
blue) (Fig.~\ref{fig2}A). In the region below -1~eV, there is a good
correspondence between them. Notice however the systematic down shift
(indicated by a green arrow) of the LDA+DMFT features relative to the
LDA bands (which have mainly In-5p and Ir-5d character). Surprisingly,
a similar trend is seen in the angle resolved photoemission
experiments (ARPES) of Ref.~\citen{Fujimori-2003} which we redraw in
Fig.~\ref{fig2}B.
The position of these bands is weakly temperature dependent and, if
warmed to room temperature, an almost rigid upward shift of 5~meV was
identified in our theoretical treatment.
Experimentally, it was not possible to resolve the momentum in $z$
direction therefore the same experimental data (which can be thought as
the average of the two paths from $\Gamma$-$X$ and $Z$-$R$) is
repeated in the two directions.

Near the Fermi level (between -0.5eV and +1eV), there are significant
discrepancies between the LDA bands (which in this region have
significant $f$ character) and the LDA+DMFT features.
The correlations treated by LDA+DMFT substantially modify the spectral
function features with $f$ content, transferring spectral weight into
the upper Hubbard band located around +3eV (white region in
Fig.~\ref{fig2}A).

Hubbard bands are excitations localized in real space, without a well
defined momentum, and therefore they show up as a blurred region of
spectral weight in momentum plot of Fig.~\ref{fig2}A. There is also a lower
Hubbard band around -2.5eV which is hardly detectable in this
figure. The reason is that it carries a very small spectral weight
which is redistributed over a broad frequency region as shown in
Fig.~\ref{fig1}B.

It is also useful to compare the LDA+DMFT Hubbard bands with those
obtained with more familiar LDA+U method. The latter method inserts a
sharp non-dispersive band around -2.5eV and significantly twists the
rests of the conduction bands.  For the purpose of describing the set
of bands below -1eV, the LDA+DMFT method, is therefore closer to the
LDA type of calculation with $f$ bands removed from the valence band.

To obtain further insights into the nature of the low energy spectra,
we show in Figs.~\ref{fig2}C and D the momentum resolved $f$
electron spectral function of Fig.\ref{fig1}B.
The two plots correspond to the low (10~K) and high (300~K)
temperature spectra, respectively. At room temperature a set of broad
and dispersive bands are seen, and should be interpreted as the $spd$
bands leaving an imprint in the $f$ electron spectral function due to
hybridization.
At low temperature, a narrow stripe of spectral weight appears at zero
frequency, which cuts the conduction bands and splits them into two
separate pieces, divided by a new hybridization gap.

The two straight non-dispersive bands at -0.3eV and +0.3eV can also be
identified in Fig.~\ref{fig2}C and are due to the spin-orbit
coupling \cite{Sekiyama}. The same splitting of the coherence peak can be identified in
the local spectra plotted in Fig.~\ref{fig1}B and was recently
observed in ARPES study\cite{Fujimori-2006}.

A detailed analysis of the zero energy stripe of spectra in
Fig.~\ref{fig2}C reveals that the low energy features correspond to
the three very narrow bands (the dispersion is of the order of 3~meV)
crossing the Fermi level. This is the origin of the large effective
mass and large specific heat of the material at low temperatures.
The low energy band structure and its temperature dependence are
theoretical predictions which can be verified experimentally in future
ARPES studies.


Optical conductivity is a very sensitive probe of the electronic
structure and has been applied to numerous heavy fermion materials\cite{Degiorgi}.
It is a technique which is largely complementary to
the photoemission, on two counts. It probes the bulk and not the
surface, and it is most sensitive to the itinerant $spd$ electrons,
rather than the $f$ electrons.

\begin{figure}
\includegraphics[width=1.0\linewidth]{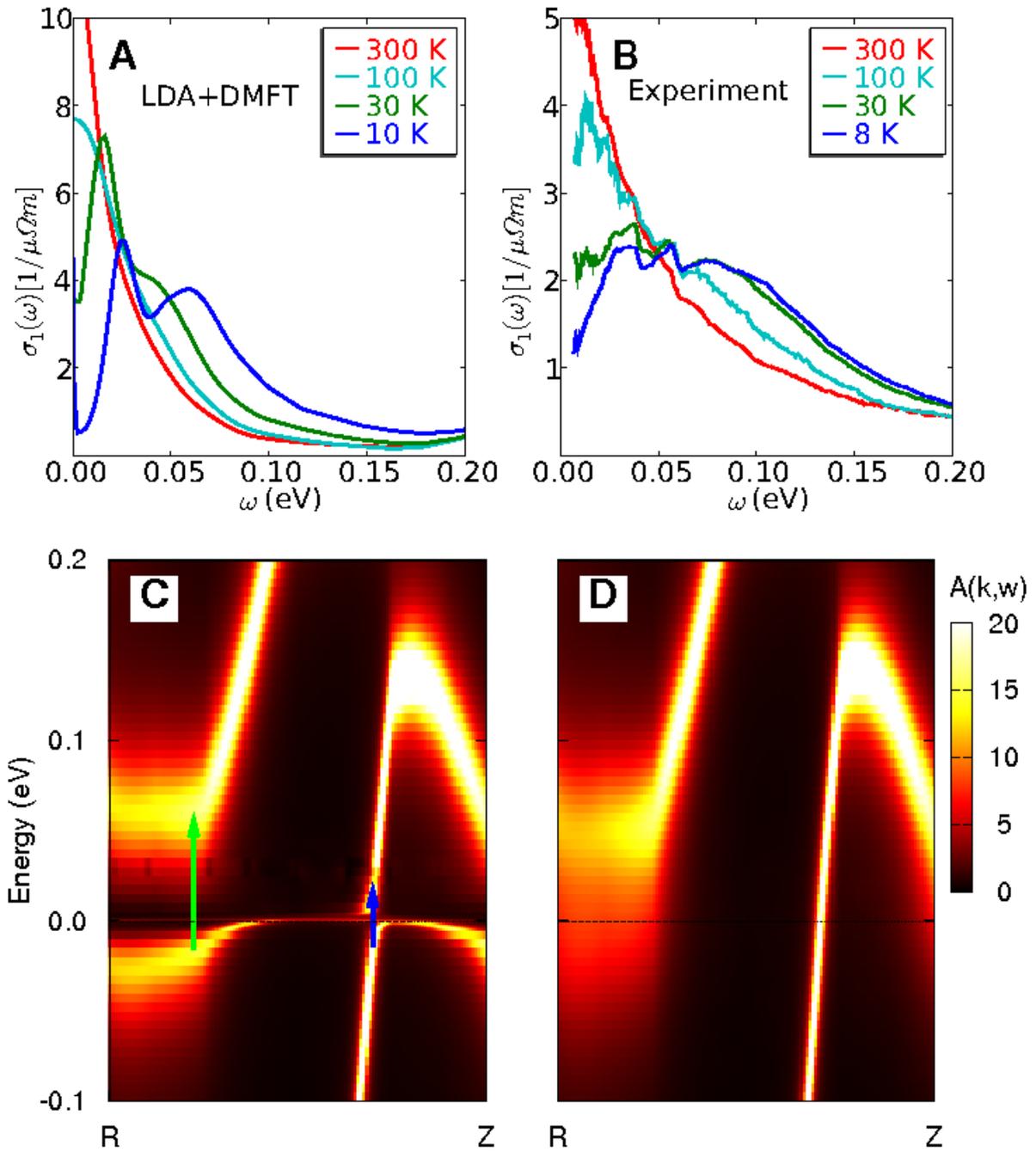}
\caption{
The optical conductivity at several temperatures
(A) obtained by LDA+DMFT and (B) measured experimentally and
reproduced from Ref.~\citen{Mena-2005}.
The momentum resolved non-f spectral function
($A_{total}$-$A_{Ce-4f}$) at (C) 10~K and (D) and 300~K.
}
\label{fig3}
\end{figure}

A prototypical heavy fermion at high temperatures has an optical
conductivity, characterized by a very broad Drude peak.  At low
temperatures, optical data is usually modeled in terms of transitions
between a two renormalized bands, separated by a hybridization
gap. These two bands give rise to a very narrow Drude peak of small
weight, and an optical absorption feature above the hybridization gap,
termed mid-infrared peak.
This picture qualitatively describes the experimental data of
CeIrIn$_5$\cite{Mena-2005,Burch}, which we reproduce in
Fig.~\ref{fig3}B.
However, this simplified two band model fails to account for some
aspects of the data.  For example, at 10~K there is a clear structure
in the mid-infrared peak.  In addition to the broad shoulder around
0.07~eV, a second peak around 0.03~eV can be identified, which was
previously interpreted as the absorption on the bosonic mode that
might bind electrons in the unconventional superconducting state\cite{Singley,Mena-2005}.

Also the hybridization gap in simplified theories gives rise to a sharp
drop of conductivity below the energy of the gap, while broader
features are seen experimentally.

Optical conductivity within LDA+DMFT was recently implemented\cite{Haule-2005}
and we show results in Fig.~\ref{fig3}A. They bear a strong
resemblance to the experimental data Fig.~\ref{fig3}B. For example
a broad Drude peak at high temperature and a very clear splitting of the
mid-infrared peak at low temperature.

To understand the physical origin of these multiple peaks we plot in
Figs.~\ref{fig3}C and \ref{fig3}D the momentum resolved conduction
electron (non-Ce-$4f$) spectral function along a representative
high-symmetry line at 10~K and 300~K, respectively.
Notice the dramatic difference between the two temperatures. At high
temperature, we see two bands in this momentum direction, one very
broad and one narrower. The dispersion of the left band in
Fig.~\ref{fig3}D is due to electron electron scattering, which
broadens the band for approximately 100meV.
The character of both bands is primarily of In-$5p$, with an important
difference. The left band comes mostly from the In atoms in the
IrIn$_2$ layer, while the right band is mostly due to In in the
CeIn$_3$ layer. The latter In atoms will be called in plane (each Ce
has 4 neighbors of this type) and the former out of plane (there are 8
nearest neighbors to Ce atom). As the temperature is lowered, the two
In bands hybridize in very different way with the Ce $4f$ moment. It
is very surprising that the in-plane In atoms hybridize less with Ce
moment leading to a small hybridization gap of the magnitude 30meV
(blue arrow in Fig.~\ref{fig3}A). The out of plane In are more coupled
to Ce moment, which leads to larger hybridization gap of the order of
70meV (green arrow in Fig.~\ref{fig1}A). The existence of multiple
hybridization gaps results in the splitting of the mid-infrared peak
in optical conductivity shown in Fig.~\ref{fig3}A.

The remarkable fact that the Ce moment is more coupled to out of plane
In than in-plane In provides a natural explanation for why these
materials are sensitive to substitution of transition metal ion Ir
with Co or Rh. Namely, the out of plane In are not only strongly
coupled to Ce but also to the transition metal ion in their immediate
neighborhood, while the in-plane In are insensitive to the
substitution.

Some of the results of the microscopic theory such as the momentum
dependent hybridization \cite{Burch} and the slow buildup of coherence
\cite{Nakatsuji} were forshadowed by earlier phenomenological
approaches.  The first principles DMFT treatment places these ideas
within a microscopic framework.



In investigating the formation of the heavy fermion
state with temperature in CeIrIn$_5$, we have shown that
incorporating local correlations on the $f$ site only, allows for a
coherent description of the evolution of the one electron spectra and
the optical conductivity with temperature. The approach, provides a
natural explanation for many surprising features observed in this
material, and makes a number of quantitative predictions for the
evolution of the spectra as a function of temperature which can be
tested by ARPES measurements currently under way.
While the single site DMFT description is sufficient in a broad region
of temperatures and parameters, cluster extensions of DMFT will be
necessary to address the quantum criticality that takes place as Ir is
replaced by Rh and Co, and the possible instabilities towards
unconventional superconductivity. While model cluster DMFT studies
seem very promising, the implementation of these methods in
conjunction with realistic electronic structure remains a challenge
for the future. Furthermore, to treat other compounds of the same
class (Ir substituted by Co or Rh), the correlations on the $3d$ or
$4d$ transition metal will require GW to treat the electronic
structure.


\end{document}